\documentclass[aps,prd,nofootinbib,superscriptaddress,onecolumn,notitlepage]{revtex4-2}
\usepackage{graphicx}
\usepackage{amssymb}
\usepackage{amsmath}
\usepackage{bm}
\usepackage{float}
\usepackage{epstopdf}
\usepackage{graphics}
\usepackage{caption}
\usepackage{subfigure}
\usepackage{epsfig}
\usepackage{color}
\usepackage{multirow,array}
\usepackage{kotex}
\usepackage{tensor}
\usepackage{hyperref}

\hypersetup{
	colorlinks=true,
	linkcolor=blue,
	filecolor=magenta,      
	urlcolor=blue,
}

\newcommand{\BH}{\text{BH}}
\newcommand{\env}{\text{env}}

\newcommand{\eff}{\text{eff}}

\newcommand{\tA}{\tilde{A}}
\newcommand{\TB}{\text{B}}

\newcommand{\tm}{\tilde{m}}

\newcommand{\tr}{\tilde{r}}

\newcommand{\tc}{\tilde{c}}

\newcommand{\tG}{\tilde{G}}
\newcommand{\tsigma}{\tilde{\sigma}}

\newcommand{\te}{\tilde{e}}

\newcommand{\tM}{\tilde{M}}
\newcommand{\tQ}{\tilde{Q}}

\newcommand{\TextH}{\text{H}}

\newcommand{\tL}{\tilde{L}}
\newcommand{\tT}{\tilde{T}}

\newcommand{\SB}{\text{SB}}
\newcommand{\GR}{\text{GR}}

\newcommand{\Teq}{\text{eq}}

\newcommand{\Max}{\text{Max}}
\newcommand{\eva}{\text{eva}}
\newcommand{\av}{\text{av}}
\newcommand{\Tot}{\text{Tot}}

\newcommand{\tk}{\tilde{k}}

\newcommand{\tepsilon}{\tilde{\epsilon}}

\newcommand{\thbar}{\tilde{\hbar}}

\begin{document}
\setcounter{page}{0}

% Title modified to reflect Geometric Shielding
\title{Black Hole Thermodynamics and Geometric Shielding in Generalized Cosmological Time}

\author{Seokcheon \surname{Lee}}
\email{skylee@skku.edu}
\affiliation{Department of Physics, Institute of Basic Science, Sungkyunkwan University, Suwon 16419, Korea}

\date[]{Received }

\begin{abstract}
The Generalized Cosmological Time (GCT) framework offers an alternative phenomenological approach to addressing the Hubble tension and the observed time dilation of Type~Ia supernovae, characterized by a background parameter $b \simeq 0.04$ and an associated cosmological scaling of fundamental constants. A key conceptual question is whether such a background evolution is compatible with the stability of local, gravitationally bound systems, in particular black holes. This work examines black hole thermodynamics within the GCT framework, focusing on the geometric compatibility between a locally static region and a time-dependent cosmological background. By matching a static interior spacetime to a GCT--FLRW exterior across a timelike boundary, it is shown that the Israel junction conditions allow for the coexistence of distinct time normalizations without introducing surface stresses. In this setting, the local interior naturally admits a unit lapse function, while the background evolution is encoded in the cosmological time gauge. The resulting separation of time normalizations implies that the effective GCT parameter governing local physics is observationally indistinguishable from $b_{\mathrm{local}} \simeq 0$. Under this geometric shielding, black hole thermodynamics reduces to its standard General Relativistic form, and the Generalized Second Law is satisfied without imposing additional constraints on the background parameter $b$. These results indicate that the empirical stability of black hole thermodynamics does not directly constrain the global GCT evolution, but instead reflects a geometric decoupling between local and cosmological time gauges. Black hole stability thus emerges as a consistency condition for geometric shielding, rather than as independent evidence for or against the underlying cosmological model.
\end{abstract}

\maketitle

%----------------------------------------------------------------------------------------
%	SECTION 1
%----------------------------------------------------------------------------------------

\section{Introduction}\label{sec1}

The standard $\Lambda$CDM model provides a successful description of the large-scale structure of the Universe, yet it continues to face persistent challenges, including the Hubble tension and reported anomalies in cosmological dipoles~\cite{Buchert:2015wwr,Perivolaropoulos:2021jda}. These tensions have motivated the exploration of alternative phenomenological frameworks that modify the interpretation of cosmic expansion without introducing additional dynamical degrees of freedom. One such approach is the Generalized Cosmological Time (GCT) framework, also referred to as the minimally extended varying speed of light (meVSL) scenario, in which the effective normalization of time evolves with the scale factor through a dimensionless parameter $b$~\cite{Lee:2020zts,Lee:2023rqv}. Recent analyses of Type~Ia supernovae have suggested that a small but nonzero value, $b \simeq 0.04$, can account for observed deviations from the standard $(1+z)$ time-dilation relation~\cite{Lee:2023ucu,Lee:2024kxa}.

A central conceptual issue in such frameworks concerns locality. If effective quantities such as the speed of light $c(t)$, Newton's constant $G(t)$, and Planck's constant $\hbar(t)$ are allowed to vary with the cosmological expansion, it is natural to ask whether this evolution extends into gravitationally bound systems. Black holes (BHs), as the most extreme realizations of strong gravity, provide a particularly sensitive arena in which to test this question. Variations of fundamental constants in the context of BH thermodynamics have been discussed previously \cite{MacGibbon:2007qh}, but a geometrically consistent treatment that reconciles a time-evolving cosmological background with the apparent stability of local gravitational physics remains incomplete. In the absence of an effective decoupling mechanism, cosmological variations would generically introduce additional contributions to entropy evolution, raising concerns regarding the validity of the Generalized Second Law (GSL).

This work examines this issue within the GCT framework by considering the geometric relationship between local and cosmological time normalizations. The analysis is based on a composite spacetime in which a locally static Schwarzschild interior is embedded in an expanding Friedmann--Lema\^{i}tre--Robertson--Walker (FLRW) exterior characterized by a time-dependent lapse function $N(t) \propto a^{b/4}$~\cite{Lee:2023xfg}. The focus is not on introducing a new dynamical screening mechanism, but on clarifying how distinct time gauges may coexist within a single spacetime description.

The key concept explored here is \textit{geometric shielding}, understood as the kinematic separation between a locally static region and a cosmological background with evolving time normalization. Unlike dynamical screening mechanisms such as the chameleon or Vainshtein effects, which rely on nonlinear interactions of additional scalar fields in high-density environments \cite{Khoury:2003aq,Khoury:2003rn,Vainshtein:1972sx}, geometric shielding arises from the matching conditions required by spacetime geometry itself. By applying the Israel junction conditions across a timelike boundary separating the two regions, it is shown that the induced metric and extrinsic curvature can remain continuous without invoking surface stresses~\cite{Israel:1966rt}. Within this construction, the local interior naturally admits a unit lapse function, while the cosmological evolution is encoded entirely in the exterior time gauge.

This separation implies that the effective parameter governing local physics is observationally indistinguishable from $b_{\mathrm{local}} \simeq 0$, even when the background evolution is characterized by $b \neq 0$. Under these conditions, black hole thermodynamics reduces to its standard General Relativistic form. The GSL is preserved without imposing additional constraints on the background evolution parameter, indicating that the stability of local gravitational systems reflects a geometric decoupling rather than a restriction on the global cosmological model.

The structure of the paper is as follows.  Section~\ref{sec:Shielding} presents the geometric basis of shielding through the Israel junction conditions and clarifies the assumptions underlying the separation between local and cosmological time normalizations. 
Section~\ref{sec:Thermo} examines black hole thermodynamics in this setting and shows that the Generalized Second Law remains consistent within a geometrically shielded framework.  The Appendix provides a detailed thermodynamic analysis of charged black holes within the meVSL parametrization, including a counterfactual comparison with a fully coupled scenario, thereby illustrating why a geometric decoupling between local and cosmological time gauges is required for consistency.

%----------------------------------------------------------------------------------------
%	SECTION 2
%----------------------------------------------------------------------------------------

\section{Geometric Shielding via Metric Junction}
\label{sec:Shielding}

In the GCT framework, the background spacetime is described by a Robertson--Walker (RW) geometry endowed with a time-dependent lapse function $N(t) \propto a^{b/4}$. This construction may be expressed as an effective cosmological scaling of quantities such as $c(t) = c_0 a^{b/4}$, $G(t) = G_0 a^{b}$, and $\hbar(t) = \hbar_0 a^{-b/4}$ \cite{Lee:2020zts,Lee:2024zcu}. While this parametrization captures deviations from the standard expansion history on cosmological scales, its extension to gravitationally bound systems requires a careful assessment of how local and global time normalizations may coexist within a single spacetime description.

To examine this issue, the Israel junction conditions are applied to match a locally static interior spacetime to a time-dependent cosmological background. A composite spacetime is considered in which an interior region $\mathcal{M}^{-}$, representing a virialized and gravitationally bound system, is embedded within an exterior region $\mathcal{M}^{+}$, representing the expanding GCT universe. The two regions are separated by a timelike hypersurface $\Sigma$, which serves as an idealized boundary between local and cosmological domains.

The interior region $\mathcal{M}^{-}$ is modeled by the standard Schwarzschild metric with a constant speed of light $c_0$ and unit lapse ($N_{-}=1$)~\cite{Lee:2023xfg}
\begin{equation}
ds_{-}^2 = -f(r)\, c_0^2 dt_{-}^2 + \frac{1}{f(r)}dr^2 + r^2 d\Omega^2 \,,
\label{eq:metric_in}
\end{equation}
where $f(r)=1-\frac{2G_0M}{c_0^2 r}$. This metric admits a timelike Killing vector and provides a well-defined notion of local energy and time consistent with standard General Relativity.

The exterior region $\mathcal{M}^{+}$ is described by an expanding GCT--FLRW geometry. Adopting the coordinate convention $x^0 = c_0 t$, the line element takes the form
\begin{equation}
ds_{+}^2 = -c(t)^2 dt_{+}^2 + a(t)^2\left(d\chi^2+\chi^2 d\Omega^2\right)\,,
\label{eq:metric_out}
\end{equation}
where $c(t)=c_0 a(t)^{b/4}$ and the associated lapse function is $N_{+}(t)=a(t)^{b/4}$.

The matching across the boundary $\Sigma$, chosen here as a comoving worldtube ($\chi=\chi_b$), requires continuity of both the induced metric and the extrinsic curvature, $[K_{ab}]=0$. Continuity of the induced metric relates the proper time $\tau$ measured on $\Sigma$ to the cosmological time coordinate $t_{+}$ according to
\begin{equation}
\frac{dt_{+}}{d\tau} = \frac{c_0}{c(t_{+})} = a(t_{+})^{-b/4}\,.
\end{equation}
This relation reflects the difference in time normalization between the local static frame and the cosmological background.

The second junction condition, enforcing continuity of the extrinsic curvature, provides an additional constraint on the evolution of the boundary. For a pressureless boundary layer, the matching yields a Friedmann-type relation governing the exterior expansion,
\begin{equation}
\frac{H^2}{c(t_{+})^2} = \frac{8\pi G_0}{3c_0^2}\rho_m \,.
\end{equation}
This expression coincides with the background equations of motion in the GCT framework and ensures that the presence of a locally static interior does not disrupt the cosmological evolution.

This result may be interpreted as a \emph{geometric consistency condition}: the existence of a locally static, virialized region characterized by fixed local scales is geometrically compatible with a cosmological background exhibiting an evolving time gauge. In this construction, the interior region admits a unit lapse and a timelike Killing vector, while the time dependence associated with the parameter $b$ is entirely encoded in the exterior geometry.

As a consequence, the effective parameter governing local physics is observationally indistinguishable from $b_{\mathrm{local}} \simeq 0$, even when the background evolution is characterized by $b \neq 0$. This geometric separation underlies the notion of \emph{geometric shielding}: local notions of time, energy, and fundamental scales remain effectively frozen within the bound system, without requiring additional dynamical screening mechanisms or fine-tuned constraints on the background evolution.

%----------------------------------------------------------------------------------------
%	SECTION 3
%----------------------------------------------------------------------------------------
\section{Robustness of Black Hole Thermodynamics}
\label{sec:Thermo}

The implications of geometric shielding for black hole thermodynamics are now examined. The Bekenstein--Hawking entropy is given by \cite{Bekenstein:1973ur}
\begin{equation}
    S_{\BH} = \frac{k_{\mathrm{B}} c^3}{\hbar G} \frac{A_{\BH}}{4} \,.
\end{equation}
In a hypothetical situation in which a black hole were fully coupled to the cosmological background evolution, such that the effective local parameter satisfied $b_{\mathrm{local}} = b_{\mathrm{background}} \simeq 0.04$, the time derivative of the entropy would acquire additional contributions proportional to $\dot{c}/c$, $\dot{G}/G$, and $\dot{\hbar}/\hbar$. These terms would introduce explicit dependencies on the Hubble expansion rate $H(t)$ and the GCT parameter $b$, modifying the standard entropy balance.

As discussed in Appendix~\ref{sec:EntBH}, such contributions can become comparable to, or even dominate over, the area variation term for certain mass and charge configurations. In these cases, maintaining the GSL would require restrictions on the sign or magnitude of $b$, which would not follow naturally from the cosmological framework itself. This sensitivity highlights a potential tension between cosmological evolution of effective constants and the observed robustness of black hole thermodynamics, if no decoupling mechanism is present.

Within the geometric shielding picture developed in the previous section, this tension is avoided. The interior of a virialized system admits a unit lapse and a timelike Killing vector, implying that local notions of time and energy remain well defined and insensitive to the background time normalization. As a result, the effective values of $c$, $G$, and $\hbar$ governing black hole physics are observationally indistinguishable from constants on local timescales, even though the cosmological background evolves.

Under these conditions, the entropy evolution is controlled solely by physical processes that change the horizon area, such as mass accretion, charge exchange, or Hawking radiation \cite{Hawking:1975vcx},
\begin{equation}
    \frac{dS_{\BH}}{dt} = \frac{k_{\mathrm{B}} c_0^3}{4 \hbar_0 G_0} \frac{dA_{\BH}}{dt} \,.
\end{equation}
This expression coincides with the standard General Relativistic result and is independent of the background GCT parameter $b$.

Since conventional black hole thermodynamics satisfies the GSL, $\Delta S_{\mathrm{total}} = \Delta S_{\BH} + \Delta S_{\env} \ge 0$, the geometrically shielded configuration inherits this property without modification. The stability of black hole thermodynamics thus does not impose additional constraints on the background cosmological evolution. Instead, it is naturally accommodated by a geometric separation between local and cosmological time gauges.

In this sense, black hole thermodynamics provides a consistency check on the geometric shielding construction rather than an independent probe of the background GCT dynamics. The absence of observable violations of the GSL in strong-gravity systems is consistent with, but does not by itself uniquely select, the presence of a geometrically shielded local time normalization.

%----------------------------------------------------------------------------------------
%	SECTION 4
%----------------------------------------------------------------------------------------
\section{Conclusion}
\label{sec:Con}

This work has examined the thermodynamic behavior of black holes within the Generalized Cosmological Time (GCT) framework, with particular attention to the compatibility between a time-dependent cosmological background and local gravitational physics. While the GCT parametrization allows for an effective cosmological evolution of quantities associated with time normalization, characterized by $b \simeq 0.04$, the implications of this evolution for strongly bound systems require careful geometric interpretation.

By applying the Israel junction conditions, it has been shown that a locally static interior spacetime can be consistently matched to an evolving GCT--FLRW background across a timelike boundary. This construction allows distinct time gauges to coexist without introducing surface stresses or disrupting the background cosmological dynamics. Within such a geometrically matched spacetime, the local interior admits a unit lapse and a timelike Killing vector, while the background evolution is encoded entirely in the exterior time normalization.

As a consequence, black hole thermodynamics within the locally static region reduces to its standard General Relativistic form. The Bekenstein--Hawking entropy and its evolution remain governed by horizon area variations alone, and the Generalized Second Law is satisfied without imposing additional constraints on the background GCT parameter. The empirical robustness of black hole thermodynamics is therefore compatible with, rather than restrictive of, the cosmological time evolution described by the GCT framework.

These results support the view that geometric shielding provides a consistent kinematic separation between local and cosmological time normalizations. Rather than constituting an independent test of the background dynamics, black hole stability serves as a consistency requirement that any viable extension of cosmological time must satisfy. In this sense, the GCT framework offers an example of how cosmological-scale modifications may coexist with the observed stability of local gravitational systems through purely geometric considerations.

%----------------------------------------------------------------------------------------
%	BIBLIOGRAPHY
%----------------------------------------------------------------------------------------

% Appendix omitted as requested. 
% Refer to previous version for "Basics of BH" and "Entropy of BH system" sections.

%----------------------------------------------------------------------------------------
%	APPENDIX
%----------------------------------------------------------------------------------------
\appendix

\section{Basics of BH}
\label{sec:BasBH} 

In this Appendix, the thermodynamic properties of charged black holes are reviewed within the meVSL (or GCT) parametrization.
The purpose of this analysis is not to describe the dynamical time evolution of a single black hole embedded in an expanding universe~\cite{McVittie:1933zz}, but to compare stationary black-hole configurations defined at different cosmological epochs labeled by the scale factor $a$. The Reissner–Nordstr\"{o}m (RN) metric is the unique static solution to the Einstein-Maxwell field equations corresponding to the gravitational field of a non-rotating spherically symmetric charged object of mass $M$ 
	\begin{align}
	&d s^2 = - \frac{c^2}{g_{rr}} dt^2 + g_{rr} dr^2 + r^2 \left( d \theta^2 +  \sin^2 \theta d \phi^2 \right) \label{RNmetric} \quad , \quad \text{where}  \\
	&g_{rr} = \left( 1 - \frac{\tr_{s}}{r} + \frac{\tr_{Q}^2}{r^2} \right)^{-1} \quad , \quad \tr_{s} = \frac{2 \tG \tM}{\tc^2} = \frac{2 \tG_0 \tM_0}{\tc_0^2}  \quad ,  \quad \tr_{Q}^2 = \frac{\tQ^2 \tG}{4 \pi \tepsilon \tc^4} = \frac{\tQ_0^2 \tG_0}{4 \pi \tepsilon_0 \tc_0^4} a^{-b/4} \label{rsrQ} \,, \\
	&\tr_{\BH} \equiv \frac{\tr_{s}}{2} \left( 1 + \sqrt{ 1 - \left( 2 \tr_{Q}/ \tr_s \right) ^2} \, \right)  \equiv \frac{\tG_0 \tM_0}{\tc_0^2} \left( 1 + \sqrt{1 -B(a) } \, \right) \,, \quad \text{where} \nonumber \\
	& B(a) = \frac{\tQ_0^2}{4 \pi \tepsilon_0 \tG_0 \tM_0^2}  a^{-b/4} \equiv B_0 a^{-b/4}  \,, \quad B_0 = 3.41 \times 10^{-41} \left( \frac{\text{M}_{\odot}}{\tM_0} \right)^2 \left( \frac{\tQ_0}{\text{C}} \right)^{2} \label{rBH} \,, \nonumber \\
	&
	\end{align}
where we use SI units and $M_{\odot} \, (C)$ denotes the solar mass (the coulomb). Also, $B(a)$ is a dimensionless quantity evolving as a function of a scale factor, $a$. 

%-------------------------------------------------
\subsection*{Local stationarity and cosmological labeling}
\label{subsec:local_stationarity}
%-------------------------------------------------

In the GCT framework, the appearance of cosmological scaling in local quantities should not be interpreted as a genuine
time evolution of a single local system. Rather, each value of the cosmological scale factor $a$ labels a distinct local
theory defined on a spacetime patch in which the local geometry and dynamics are time-independent.

In particular, relations such as
$\tilde M(a)=\tilde M_0 a^{-b/2}$ and $\tilde Q(a)=\tilde Q_0 a^{-b/4}$
do not describe the temporal evolution of the mass or charge of an individual
black hole.
For any fixed value $a=a_1$, the corresponding Reissner--Nordström solution is
exactly static, with $\tilde M(a_1)$ and $\tilde Q(a_1)$ treated as constants
characterizing that local spacetime.

The dependence on $a$ enters only when comparing stationary black-hole systems
defined at different cosmological epochs.
This structure is conceptually analogous to the renormalization-group description
in quantum field theory, where coupling constants depend on an external scale
parameter while each theory at fixed scale remains autonomous and stationary.

Accordingly, throughout this Appendix, derivatives with respect to time should be
understood as parametrizing comparisons between neighboring cosmological slices
labeled by $a(t)$, rather than as describing local time evolution within a single
black-hole spacetime.
No assumption is made regarding a continuous deformation of the Reissner--Nordström
geometry inside a locally stationary region.

\begin{figure*}
\centering
\vspace{1cm}
\begin{tabular}{cc}
\includegraphics[width=0.5\linewidth]{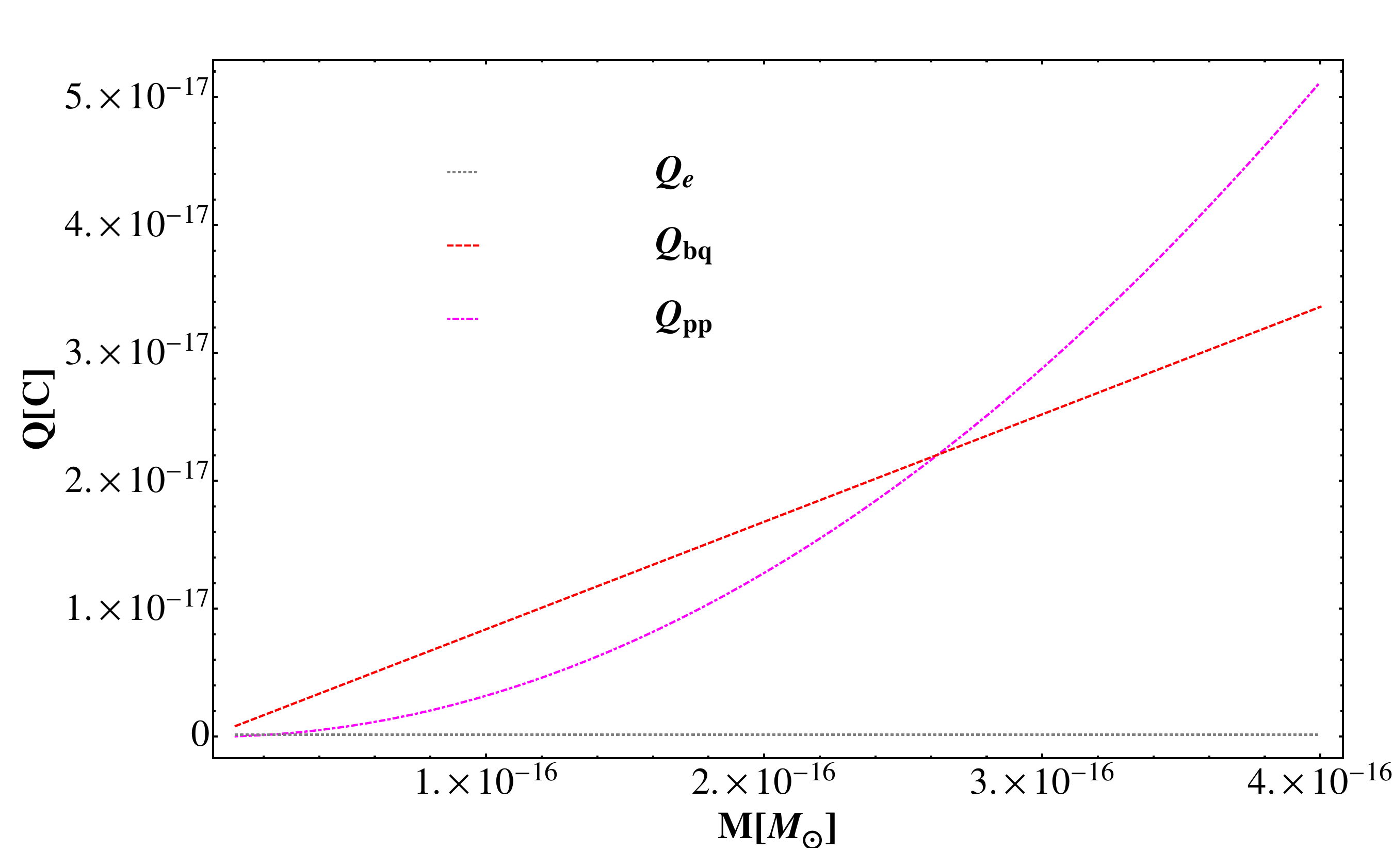} &
\includegraphics[width=0.49\linewidth]{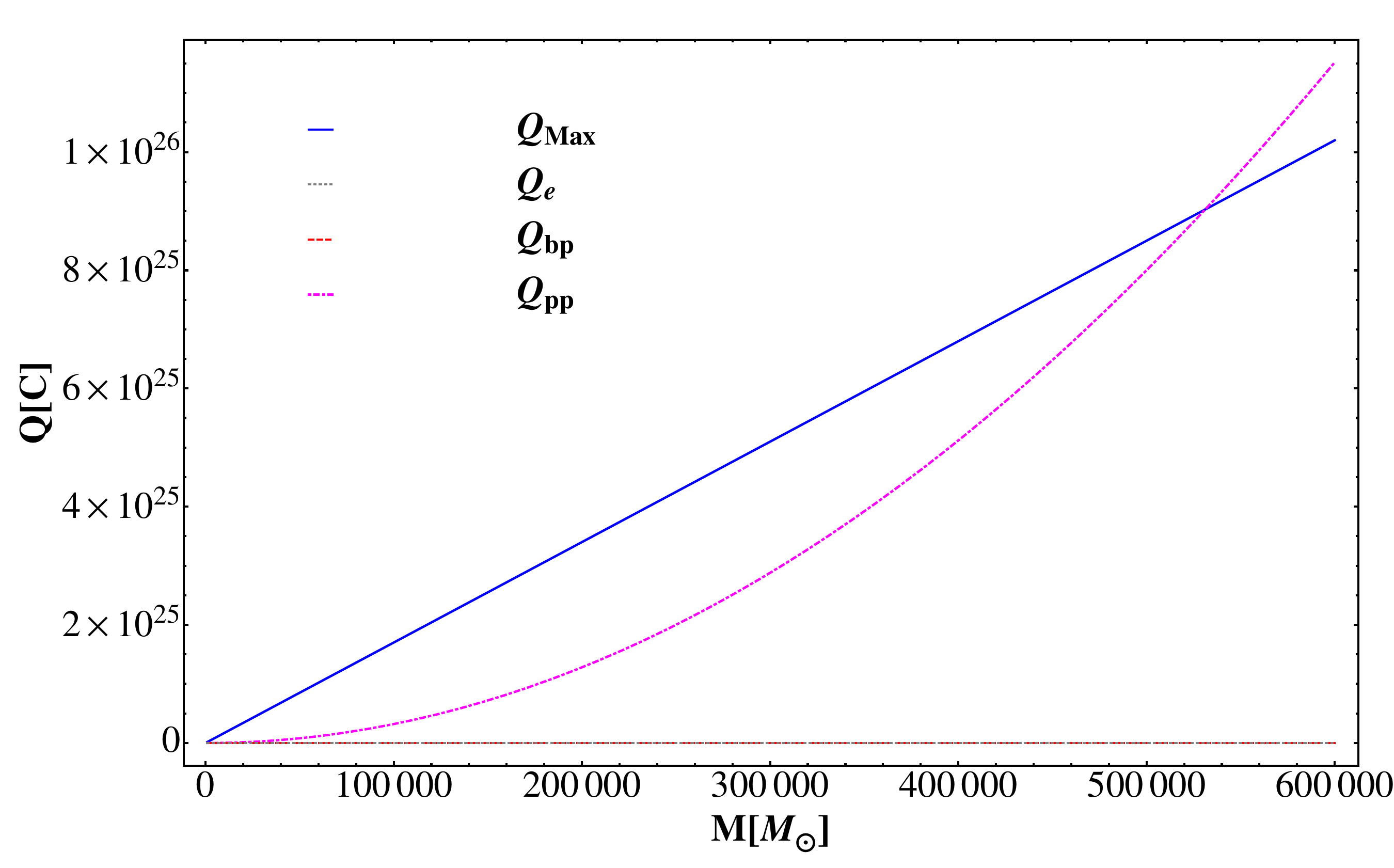}
\end{tabular}
\vspace{-0.5cm}
\caption{The behaviors of different definitions of BH charges as a function of BH mass. The solid, dotted, dashed, and dot-dashed lines represent $Q_{\Max} \,, Q_{e} \,, Q_{\text{bp}}$ \,, and $Q_{\text{pp}}$, respectively. a) The behaviors of them for the mass range up to $4 \times 10^{-16} M_{\odot}$. b) Their behaviors for the mass range to $6 \times 10^{5} M_{\odot}$.} \label{fig-1}
\vspace{1cm}
\end{figure*}

We regard time evolutions of both physical quantities (mass and charge) and physical constants (the vacuum permittivity and the Gravitational constant) of meVSL model as \cite{Lee:2020zts}
	\begin{align}
	\tM = \tM_0 a^{-b/2} \quad , \quad \tQ = \tQ_0 a^{-b/4} \quad , \quad \tepsilon = \tepsilon_0 a^{-b/4}  \quad , \quad \tG = \tG_{0} a^{b} \label{tMtQ} \,,
	\end{align}
where a subscript $0$ on each quantity implies its value at the present epoch. Thus, the Schwarzschild radius of meVSL model, $\tr_s$, is the same as that of GR. However, the characteristic length scale, $\tr_Q$, evolves as a function of scale factor, $a$. There exists the condition between them, $2 r_{Q} \leq r_{s}$, in order to have the physical event horizon as shown in Eq.~\eqref{rBH}. Objects with $2r_{Q} > r_{s}$ can exist but they are not able to collapse down to form a BH. Thus, the maximum charge of a BH is given by
\begin{align}
	\tQ_{\Max} = \sqrt{4 \pi \tepsilon_0 \tG_0} \tM_0 a^{b/8} = 1.71 \times 10^{20} \left( \frac{\tM_{ 0}}{M_{\odot}} \right) a^{b/8} [C] \label{Qmax} \,.
\end{align}
However, it has been shown that large BHs should neither acquire significant charge nor approach $\tQ_{\Max}$ \cite{Gibbons:1975kk}. Also, one expects that the density distributions of electrons and protons are comparable due to the quasineutrality of BH \cite{Bally:1978apj, Zajacek:2019kla}. This leads to the equality of potential of electrons and protons and makes us obtain a value of the equilibrium charge 
\begin{align}
\tQ_{\Teq}  = \frac{2 \pi \tepsilon \tG \left( \tm_{p} - \tm_{e} \right)}{\te} \tM = \frac{2 \pi \tepsilon_0 \tG_0 \left( \tm_{p 0} - \tm_{e 0} \right)}{\te_0} \tM_0 = 0.77 \times 10^{2} \left( \frac{\tM_0}{M_{\odot}} \right) [C] \label{QeqoM} \,,
\end{align}
There exists another characteristic charge scale, $Q_{\text{pp}}$, above which the BH quickly discharges by the superradiant Schwinger-type $e^{+} e^{-}$ pair-production in the electrostatic field surrounding the BH
\begin{align}
	\tQ_{\text{pp}} &= 4 \pi \tepsilon \frac{\tG^2 \tm_e^2 \tM^2}{\thbar \tc \te} = 4 \pi \tepsilon_0 \frac{\tG_0^2 \tm_{e0}^2 \tM_0^2}{\thbar_0 \tc_0 \te_0} = 3.23 \times 10^{14} \left( \frac{\tM_0}{M_{\odot}} \right)^2 \left[C \right] \label{Qpp} \,. \\
\end{align}
Another interesting charge scale, $Q_{\text{bp}}$, above which it is energetically favorable to form pairs
\begin{align}
	\tQ_{\text{bp}} &= 4 \pi \tepsilon \frac{\tG \tm_e \tM}{\te} =  4 \pi \tepsilon_0 \frac{\tG_0 \tm_{e0} \tM_0}{\te_0} =  8.40 \times 10^{-2}  \left( \frac{\tM_0}{M_{\odot}} \right) \left[C \right] \label{Qbp} \,.
\end{align}
The mass dependences on the different definitions of charges of BH are shown in the figure.~\ref{fig-1}. The behaviors of $Q_{\Max} \,, Q_{e} \,, Q_{\text{bp}}$ \,, and $Q_{\text{pp}}$ are represented as the solid, dotted, dashed, and dot-dashed lines, respectively.   In the left panel of Fig.~\ref{fig-1}, one can find that $Q_{\text{bp}}$ can be greater than $Q_{\text{pp}}$ for the small mass of BHs. $Q_{\Max}$ should be greater than other definitions of charges. This condition cannot be satisfied for SMBHs as shown in the right panel of Fig.~\ref{fig-1}. 

%%%%%%%%%%%%%%%%%%%%%%%%%%%%%%%%%%%%%%%
%%%%%%%%%%%%%%%%%%%%%%%%%%%%%%%%%%%%%%%
Hawking has shown that a BH is continuously emitting quasi-thermal radiations with a temperature
	\begin{align}
	\tT_{\BH} &= \frac{2 \thbar \tG \tM}{\tk_{\TB} \tc \tA_{\BH}} \sqrt{1 - \frac{\tQ^2}{4 \pi \tepsilon \tG \tM^2}} = \frac{2 \thbar_0 \tG_0 \tM_0}{\tk_{\TB 0} \tc_0 \tA_{\BH}} \sqrt{1 - B(a) } %\nonumber \\ 
	= \frac{\thbar_0 \tc_0^3}{2 \pi \tk_{\TB 0} \tG_0 \tM_0} \frac{\sqrt{1-B(a)}}{\left( 1 + \sqrt{1 - B(a)} \right)^2}  \,. \label{TBH}
	\end{align}
As expected, the temperature of RN BH recovers the Schwarzschild one when $Q = 0$. The Hawking luminosity $\tL_{\BH}$ of the BH is given by the usual Stefan-Boltzmann blackbody formula
\begin{align}
\tL_{\BH} = \tA_{\BH} \tsigma_{\SB} \tT_{\BH}^{4} = \frac{\thbar_0 \tc_0^6}{\tG_0^2 \tM_0^2} a^{b/4} \frac{\left(1-B(a)\right)^2}{240 \pi \left( 1 + \sqrt{1 - B(a)} \right)^6} \label{LBH} \,,
\end{align}	
where we use the Stefan-Boltzmann constant
\begin{align}
\tsigma_{\SB} \equiv \frac{\pi^2}{60} \frac{\tk_{\TB}^4}{\tc^2 \thbar^3} = \frac{\pi^2}{60} \frac{\tk_{\TB 0}^4}{\tc_{0}^2 \thbar_{0}^3} a^{b/4} \equiv \tsigma_{\SB 0} a^{b/4} \,.  \label{tsigmaSB}
\end{align} 
If the Hawking temperature exceeds the rest mass energy of a particle, then the BH radiates particles and antiparticles in addition to photons. Luminosity, $\tL_{\BH}$, given in Eq.~\eqref{LBH}, actually means the energy output and thus
\begin{align}
\frac{d \tilde{E}_{\BH}}{dt} \equiv -\tL_{\BH} = -  \frac{\thbar_0 \tc_0^6}{\tG_0^2 \tM_0^2} a^{b/4} \frac{\left(1-B(a)\right)^2}{240 \pi \left( 1 + \sqrt{1 - B(a)} \right)^6} \label{dEdt} \,.
\end{align}
From the above equation \eqref{dEdt}, one can obtain the mass loss due to Hawking radiation as 
\begin{align}
\frac{dM_{\TextH}}{dt} &\equiv \frac{1}{\tc^2} \frac{d \tilde{E}_{\BH}}{dt} \approx  -  \frac{\thbar_0 \tc_0^4}{\tG_0^2 \tM_0^2} a^{-b/4} \frac{\left(1-B(a)\right)^2}{240 \pi \left( 1 + \sqrt{1 - B(a)} \right)^6} \equiv - \frac{\thbar_0 \tc_0^4}{\tG_0^2 \tM_0^2} \beta(a) \label{dMdt} \quad , \quad \text{where} \\
\frac{\thbar_0 \tc_0^4}{\tG_0^2 \tM_0^2} &= 2.430 \times 10^{-71} \left( \frac{M_{\odot}}{\tM_0} \right)^2 \left[ \frac{\text{M}_{\odot}}{\text{s}} \right] \quad \text{and} \quad \beta(a) = a^{-b/4} \frac{\left(1-B(a)\right)^2}{240 \pi \left( 1 + \sqrt{1 - B(a)} \right)^6} \nonumber \,.
\end{align}
$\beta(a_0)$, the present value of $\beta$ without a charge, gives the standard value $1/(15360 \pi) \approx 2.1 \times 10^{-5}$. This mass loss rate applies only if the charge of BH is much smaller than the maximal possible charge on a BH, $\tQ_{\text{MAX}} = \sqrt{\tG_0 \tM_0^2/(4 \pi \tepsilon_0)} a^{b/8}$. Strictly, the mass loss rate in Eq.~\eqref{dMdt} applies only if $\tQ \ll \tQ_{\Max}$ and this is satisfied in this case because high $\tQ$ is quickly discharged if $\tM \leq \thbar \tc \te /(\tG^{3/2} \tm_{e}^2 \sqrt{4 \pi \tepsilon}) \approx 5 \times 10^{5}  a^{-9b/8} \left[ M_{\odot} \right]$. 

We compare the values of these charges for different (mass) types of BHs in table~\ref{tab:BHsMQT}. The MBH denotes the micro BH which has the Planck mass, $M \sim 10^{-38} M_{\odot}$. We define YBH as the youngest BH which has the minimum mass ($M \sim 10^{-19} M_{\odot}$) to exist at present. Also, we define the BH with an electron mass as the specific BH. SMBH denotes the super massive BH. We also show both temperatures ($T_{\BH}$) and the evaporation times ($t_{\eva}$) for the different BHs in this table. Values of $Q_{\text{pp}}$ of SMBHs are larger than those of $Q_{\Max}$ and thus it is impossible to obtain $Q_{\text{pp}}$ for SMBHs. Usually, $Q_{\text{pp}} \gg Q_{\text{bp}}$ for astrophysical BHs but this relation is opposite for the smaller BHs ({\it i.e.}, $\text{M} < 10^{-17} M_{\odot}$) as shown in table~\ref{tab:BHsMQT}.
\begin{table}[htbp]
	\centering
\begin{tabular}{|c|c|c|c|c|c|c|}
	\hline 
	& MBH & YBH & specific BH  & stellar mass BH & intermediate BH & SMBH \\
	\hline
	M [$M_{\odot}$] & $10^{-38}$ & $10^{-19}$ & $10^{-17}$ & $4 \sim 10^{2}$ & $10^{2} \sim 10^{5}$ & $10^{5} \sim 10^{10}$ \\
	\hline
	$Q_{\Max}^{0}$ [$C$] & $10^{-18}$ & $10$ & $10^{3}$ & $10^{20} \sim 10^{22}$ & $10^{22} \sim 10^{25}$ & $10^{25} \sim 10^{30}$ \\
	\hline
	$Q_{\text{pp}}$ [$C$] & $10^{-62}$ & $10^{-24}$ & $10^{-20}$ & $10^{14} \sim 10^{18}$ & $10^{18} \sim 10^{24}$ & $10^{24} \sim 10^{34}$ \\
	\hline
	$Q_{\Teq}$ [$C$] & $10^{-36}$ & $10^{-17}$ & $10^{-15}$ & $10^{2} \sim 10^{4}$ & $10^{4} \sim 10^{7}$ & $10^{7} \sim 10^{12}$ \\
	\hline
	$Q_{\text{bp}}$ [$C$] & $10^{-40}$ & $10^{-21}$ & $10^{-19}$ & $10^{-2} \sim 10^{0}$ & $10^{0} \sim 10^{3}$ & $10^{3} \sim 10^{8}$ \\
	\hline
	$t_{\eva}$ [yrs] & $10^{-47}$ & $10^{10}$ & $10^{16}$ & $10^{68} \sim 10^{73}$ & $10^{73} \sim 10^{82}$ & $10^{82} \sim 10^{97}$ \\
	\hline
	$T_{\BH}$ [$K$] & $10^{30}$ & $10^{11}$ & $10^{9}$ & $10^{-8} \sim 10^{-10}$ & $10^{-10} \sim 10^{-13}$ & $10^{-13} \sim 10^{-18}$ \\
	\hline
	$k_{\text{B}} T_{\BH}$ [$eV$] & $10^{26}$ & $10^{7}$ & $10^{5}$ & $10^{-12} \sim 10^{-14}$ & $10^{-14} \sim 10^{-17}$ & $10^{-17} \sim 10^{-22}$ \\
	\hline
\end{tabular}
\caption{The different definitions of charges, the evaporation times, and the temperatures for different BHs. MBH denotes micro BH and YBH means the youngest BH which has the evaporation time equal to the age of the Universe. The specific BH means that its thermal energy is the same as the rest mass energy of the electron. SMBH denotes a supermassive black hole. Also, yrs means years.}
\label{tab:BHsMQT}
\end{table}

\section{Entropy of BH system}
\label{sec:EntBH} 

The entropy of a BH is given by \cite{Hawking:1975vcx,Hawking:1974rv}
	 \begin{align}
	S_{\BH} = \frac{\tk_{\TB} \tc^3}{\thbar \tG} \frac{\tA_{\BH}}{4} = \frac{k_{\TB 0} \tc_0^3}{\thbar_0 \tG_0} \frac{\tA_{\BH}}{4} \quad , \quad \tA_{\BH} = 4 \pi \tr_{\BH}^2 \label{SBH} \,, 
	\end{align} 
where we rewrite it for the meVSL model and $\tA_{\BH}$ is the area of a BH's event horizon of the meVSL model. Time evolutions of quantities as functions of a scale factor, $a$, are given by \cite{Lee:2020zts}
	\begin{align}
	\tk_{\TB} = \tk_{\TB 0} \quad , \quad \tc = \tc_0 a^{b/4} \quad , \quad \thbar = \thbar_{0} a^{-b/4} \label{tquantities} \,,
        \end{align}
where $k_{\TB}$ is the Boltzmann constant, $c$ is the speed of light, and $\hbar$ is the reduced Planck constant. Thus, the prefactor of the BH entropy of meVSL model, $\tk_{B} \tc^3/(\thbar \tG)$ is the same as that of GR even though the physical constants are functions of a scale factor as shown in Eq.~\eqref{SBH}. 
The generalized second law of thermodynamics of the black hole (BH) system states that the net entropy of the system cannot decrease with time \cite{Bekenstein:1973ur}. From this fact, we might be able to constrain time variations of physical constants \cite{MacGibbon:2007qh}. The increase of the net generalized entropy of the system over a time interval $\Delta t$ is given by
	\begin{align}
	\Delta S \equiv \Delta S_{\BH} + \Delta S_{\env} \geq 0 \label{ent1} \,, 
	\end{align} 
where $\Delta S_{\BH}$ and $\Delta S_{\env}$ denote the changes in the entropy of the BH and of the environment (the ambient radiation and/or matter), respectively. From its definition given in Eq.~\eqref{SBH}, the change in BH entropy over time should include the contribution from the Hawking flux as well as partial change induced by time evolutions of physical constants
	\begin{align}
	 \frac{\Delta S_{\BH}^{\eff}}{\Delta t} &\approx \frac{d S_{\BH}}{dt} = \frac{k_{\TB 0} \tc_0^3}{4 \thbar_0 \tG_0} \frac{d \tA_{\BH}}{dt} \quad , \quad \text{where} \nonumber \\
	 \frac{d \tA_{\BH}}{dt} &= \frac{8 \pi \tG_0^2 \tM_0^2}{\tc_0^4} \frac{1 + \sqrt{1 - B(a)}}{\sqrt{1 - B(a)}} \left[ \left( 1 + \sqrt{1 - B(a)} \right) \frac{1}{M} \frac{dM_{\TextH}}{dt} - B(a) \frac{1}{Q} \frac{dQ_{\TextH}}{dt} + \frac{b}{8} H B(a) \right] \label{dABHdt} \,. \\
	\frac{d S_{\BH}}{dt} &= \frac{2 \pi k_{\TB 0} \tG_0 \tM_0^2}{\thbar_0 \tc_0} \frac{1 + \sqrt{1 - B(a)}}{\sqrt{1 - B(a)}} \left[ \left( 1 + \sqrt{1 - B(a)} \right) \frac{1}{M} \frac{dM_{\TextH}}{dt} - B(a) \frac{1}{Q} \frac{dQ_{\TextH}}{dt} + \frac{b}{8} H B(a) \right] \label{dSBHdt} \\
	&\equiv  \frac{d S_{\BH}}{dt} \Bigl|_{\mathrm{I}} +  \frac{d S_{\BH}}{dt} \Bigl|_{\mathrm{II}} +  \frac{d S_{\BH}}{dt} \Bigl|_{\mathrm{III}} \nonumber \,,
	\end{align}
where both $M$ and $Q$ change as the BH radiates and this effect is described by the subscript H. Also, the effect of meVSL on the entropy change is shown in the last term of Eq.~\eqref{dSBHdt}. This term is further simplified for the late time $\Lambda$CDM cosmology as 
\begin{align}
	\frac{d S_{\BH}}{dt} \Bigl|_{\mathrm{III}} &= \frac{\tk_{\TB 0} Q^2 H_0^{(\GR)}}{16 \tepsilon_0 \thbar_0 \tc_0} b \frac{ 1 + \sqrt{1 - B}}{\sqrt{1-B}} E(a) a^{-b/4}  \approx 2.01 \times 10^{-5} \left( \frac{Q}{C} \right)^2 h \cdot b \cdot E(a) a^{-b/4} \left[ \frac{\text{J}}{\text{K} \cdot \text{s}} \right] \,, \nonumber \\
	&\text{where} \quad E(a) = \sqrt{\Omega_{m0} a^{-3} + \Omega_{\Lambda}} \label{dSBHdtIII} \,,
\end{align}
where $h$ comes from the Hubble parameter ($H_0 = 100h \, \text{km/s/Mpc}$). Here, the derivative with respect to $t$ should be understood as a bookkeeping device that parametrizes the comparison between neighboring cosmological epochs labeled by $a(t)$.
It does not correspond to a physical time derivative measured by a local observer inside a stationary black-hole spacetime. Thus, this effect increases (decreases) the entropy change in addition to the BH entropy change due to Hawking radiation if the sign of $b$ is positive (negative). This term is significant to probe the thermodynamics of BH and thus we need to investigate it in detail. In the above Eq.~\eqref{dSBHdtIII}, we use the following relation
	\begin{align}
	H B(a) &= H_{0}^{(\GR)} B_0 E(a) =1.10 \times 10^{-58} \frac{\text{M}_{\odot}^2}{\text{C}^2} \left( \frac{\tQ_0}{\tM_0} \right)^{2} h \cdot E(a) \left[ \text{s}^{-1} \right] \label{HBa} \,.
	\end{align}
In the following subsections, we compare the magnitude of each term in Eq.~\eqref{dSBHdt} for different cases.

For the emission of $n_s$ species of two-polarization massless particles of spin $s = 1/2, 1,$ and $2$ from a Schwarzschild BH into empty space, numerical calculations have been obtained \cite{Page:1976df,Page:1976ki,Page:1983ug}
\begin{align}
\frac{d S_{\text{rad}}}{dt} &= 2.8 \times 10^{-21} \left( 3.3710 n_{1/2} + 1.2684 n_1 + 0.1300 n_2  \right) \left( \frac{M_{\odot}}{M} \right) \left[ \frac{J}{K \cdot s} \right] \label{dSraddtn} \,, \\
\frac{d S_{\BH}}{dt} &= -2.8 \times 10^{-21} \left( 2.0566 n_{1/2} + 0.8454 n_1 + 0.0964 n_2  \right)  \left( \frac{M_{\odot}}{M} \right) \left[ \frac{J}{K \cdot s} \right] \label{dSBHdtn} \,.
\end{align}
Now we investigate two cases by comparing the temperature of BH with that of its environment. 

\subsection{$\tT_{\BH} > T_{\env}$}
\label{subsec:TBHgTenv} 
When the temperature of the BH is greater than that of its surroundings, there is a net radiation loss from the BH into its environment. This case also can be divided into whether the charge of the BH, $Q$, is affected by the Hawking radiation or not. 

\subsubsection{$dQ_{\TextH}/dt = 0$ ({\it i.e.}, $k_{\TB} T_{\BH} \leq m_{e} c^2$)}
\label{subsubsec:dQ0} 
This is the case when the thermal energy of the BH with its temperature given in Eq.~\eqref{TBH} is below the rest mass energy of the electron ({\it i.e.}, $0.511$ MeV). This corresponds to 
\begin{align}
\tM_{\BH} \geq 4.15 \times 10^{-17} \frac{\sqrt{1 - B(a)}}{\left( 1 + \sqrt{1 - B(a)} \right)^2} \left[ M_{\odot} \right] \label{MBH} \,.
\end{align}
Thus, one can simplify the equation~\eqref{dSBHdt} as
\begin{align}
\frac{d S_{\BH}}{dt} = \frac{d S_{\BH}}{dt} \Bigl|_{\mathrm{I}} +  \frac{d S_{\BH}}{dt} \Bigl|_{\mathrm{III}}  \label{dSBHdt1} \,,
\end{align}
One needs to adopt the mass loss due to the Hawking radiation given in the equation \eqref{dMdt}. As mentioned before, this mass loss rate in Eq.~\eqref{dMdt} applies only if $Q \ll Q_{\Max}$ and this is satisfied in this case because high $Q$ is quickly discharged if $M \leq \hbar c e /(G^{3/2} m_{e}^2 \sqrt{4 \pi \epsilon_0}) \approx 5 \times 10^{5} M_{\odot}$.  This case is limited below the mass of SMBHs. Thus, the first term in Eq.~\eqref{dSBHdt1} is rewritten as
\begin{align}
\frac{dS_{\BH}}{dt} \Bigl |_{\mathrm{I}} &= -2.34 \times 10^{-20}  \left( \frac{M_{\odot}}{\tM_0} \right) \frac{\left( 1 - B\right)^{3/2}}{\left( 1 + \sqrt{1-B} \right)^4} a^{b/4} \left[ \frac{J}{K \cdot s}\right] \label{dSBHdt1case1} \,.
\end{align}
We can also estimate the second term in Eq.~\eqref{dSBHdt1} 
\begin{align}
\frac{dS_{\BH}}{dt} \Bigl |_{\mathrm{III}} &= 2.01 \times 10^{-5} h \cdot b \left( \frac{\tQ_0}{C} \right)^2 E(a) \label{dSBHdt3case1} \,.
\end{align}

If the charge is greater than $Q_{\text{pp}}$, then the BH is quickly discharged by superradiant pair production in the electrostatic field surrounding the BH. The discharge rate for $M \geq 10^{-17} M_{\odot}$ BH with $Q > Q_{\text{pp}}$ is given by 
\begin{align}
\frac{d Q_{\text{pp}}}{dt} &\approx - \frac{e^4 Q^3}{\left( 4 \pi \epsilon \right)^3 \hbar^3 c^2 r_{\BH}} \exp \left[ - \frac{\pi m_e^2 c^3 \left( 4 \pi \epsilon \right) r_{\BH}^2}{\hbar Q e} \right] \nonumber \\ 
	&=  - \frac{\te_0^4 \tQ_0^3}{\left( 4 \pi \tepsilon_0 \right)^3 \hbar_0^3 \tG_0 \tM_0} \exp \left[ - \frac{\pi \tm_{e 0}^2  \left( 4 \pi \tepsilon_0 \right) \tG_0^2 \tM_0^2}{\thbar_0 \tc_0 \tQ_0 \te_0} a^{b/4} \left( 1 + \sqrt{1 - B} \right)^2 \right]  \frac{a^{-3b/4}}{\left( 1 + \sqrt{1 - B} \right)} \label{dQrapidodt} \,.
\end{align}
Thus, from this charge contribution, we have a new term  $(dS_{\BH}/dt)_{\mathrm{II}}$ in this case
\begin{align}
\frac{dS_{\BH}}{dt} \Bigl |_{\mathrm{II}}^{(\text{pp})} &= \frac{2 \pi \tk_{\TB 0} \te_0^4 \tQ_0^4}{\left( 4 \pi \tepsilon_0 \right)^4 \thbar_0^4 \tG_0 \tM_0 \tc_0} \exp \left[ - \frac{\pi \tm_{e 0}^2  \left( 4 \pi \tepsilon_0 \right) \tG_0^2 \tM_0^2}{\thbar_0 \tc_0 \tQ_0 \te_0} a^{b/4} \left( 1 + \sqrt{1 - B} \right)^2 \right]  \frac{a^{-3b/4}}{ \sqrt{1 - B} } \nonumber \\
	&= 7.68 \times 10^{49} \left( \frac{\tQ_0}{C} \right)^4 \left( \frac{M_{\odot}}{\tM_0} \right) \exp \left[ - 1.01 \times 10^{15} \left( \frac{C}{\tQ_0} \right)  \left(\frac{\tM_0}{M_{\odot}} \right)^2  \right] \left[ \frac{J}{K \cdot s} \right] \label{dSBHdtIIpp} \,.
\end{align}

As shown in Eqs.~\eqref{dSraddtn} and \eqref{dSBHdtn}, the increase of the entropy of the environment due to the Hawking radiation is $1.62$ times bigger than the decrease of the entropy of BH due to Hawking emission ($dS_{\env}/dt = 1.62 dS_{\BH}/dt |_{\mathrm{I}}$). Thus, the total entropy of the system can be divided into two cases for different values of charge
\begin{align}
\frac{dS_{\text{tot}}}{dt} = \frac{dS_{\env}}{dt} +  \frac{dS_{\BH}}{dt} &= \begin{cases} 1.62 \frac{dS_{\BH}}{dt}  \Bigl|_{\mathrm{I}}  - \frac{dS_{\BH}}{dt}  \Bigl|_{\mathrm{I}}  + \frac{dS_{\BH}}{dt} \Bigl|_{\mathrm{II}}^{(\text{pp})} +  \frac{dS_{\BH}}{dt} \Bigl|_{\mathrm{III}}^{(\text{pp})} \equiv \frac{dS_{\text{tot}}^{(\text{pp})}}{dt}  & \,,  Q_{\text{pp}} < Q \, \\ 1.62 \frac{dS_{\BH}}{dt}  \Bigl|_{\mathrm{I}}  - \frac{dS_{\BH}}{dt}  \Bigl|_{\mathrm{I}}  +  \frac{dS_{\BH}}{dt} \Bigl|_{\mathrm{III}}^{(\text{eq})}  \equiv \frac{dS_{\text{tot}}^{(\text{eq})}}{dt}& \,, Q_{\text{eq}} < Q  < Q_{\text{pp}} \,
 \end{cases}  \label{gsl1} \,,
\end{align}
One can estimate each term of the above Eq.~\eqref{gsl1} for the effective mass range of BHs. We show this in Table~\ref{tab:SBHcaseI}. 
\begin{table}[htbp]
	\centering
\begin{tabular}{|c|c|c|c|}
	\hline 
	& specific  & stellar mass BH & intermediate BH \\
	\hline
	M [$M_{\odot}$] & $10^{-17}$ & $10^{0} \sim 10^{2}$ & $10^{2} \sim 10^{5}$  \\
	\hline
	$Q_{\text{pp}}$ [$C$] & $10^{-20}$ & $10^{14} \sim 10^{18}$ & $10^{18} \sim 10^{24}$ \\
	\hline
	$Q_{\Teq}$ [$C$] & $10^{-15}$ & $10^{2} \sim 10^{4}$ & $10^{4} \sim 10^{7}$  \\
	\hline
	$dS_{\BH}/dt|_{\mathrm{I}} $ [$J/(K\cdot s)$] & $10^{-4}$ & $10^{-21} \sim 10^{-23}$ & $10^{-23} \sim 10^{-26}$  \\
	\hline
	$dS_{\BH}/dt|_{\mathrm{II}}^{(\text{pp})} $ [$J/(K\cdot s)$] & $10^{-14}$ & $10^{105} \sim 10^{120}$ & $10^{120} \sim 10^{140}$  \\
	\hline
	$dS_{\BH}/dt|_{\mathrm{III}}^{(\text{pp})}$ [$J/(K\cdot s)$] & $10^{-45} (h\cdot b)$ & $10^{23} (h\cdot b) \sim 10^{31}(h\cdot b) $ & $10^{31}(h\cdot b) \sim 10^{43}(h\cdot b)$  \\
	\hline
	$dS_{\BH}/dt|_{\mathrm{III}}^{(\text{eq})} $ [$J/(K\cdot s)$] & $10^{-35} (h\cdot b)$ & $10^{-1} (h\cdot b) \sim 10^{3} (h\cdot b)$ & $10^{3} (h\cdot b) \sim 10^{9} (h\cdot b)$  \\
	\hline
	$dS_{\text{tot}}^{(\text{pp})}/dt $ [$J/(K\cdot s)$] & $10^{-4}$ & $10^{105} \sim 10^{120}$ & $10^{120} \sim 10^{140}$  \\
	\hline
	$dS_{\text{tot}}^{(\text{eq})}/dt $ [$J/(K\cdot s)$] & $10^{-4}$ & $10^{-1} (h\cdot b) \sim 10^{3} (h\cdot b)$ & $10^{3} (h\cdot b) \sim 10^{9} (h\cdot b)$  \\
	\hline
\end{tabular}
\caption{Comparisons of entropy changes from different contributions for the different BHs.}
\label{tab:SBHcaseI}
\end{table}
As shown in this table, the term $dS_{\BH}/dt|_{\mathrm{I}}$ dominates when $\tM_0 = 10^{-17} M_{\odot}$, independently of the magnitude of the black-hole charge, so that the total entropy increases regardless of the sign of $b$. For black holes with masses in the range $1$--$10^{5} M_{\odot}$, the contribution $dS_{\BH}/dt|_{\mathrm{II}}^{(\text{pp})}$ dominates, and the generalized entropy again increases for either sign of $b$. 

In contrast, when the black-hole charge is taken to be $Q_{\text{eq}}$, the entropy balance becomes dominated by the term $dS_{\BH}/dt|_{\mathrm{III}}^{(\text{eq})}$. In this hypothetical fully coupled scenario, the sign of $b$ can influence the entropy balance for specific charge configurations. This sensitivity illustrates that, in the absence of a decoupling mechanism, black-hole thermodynamics may become dependent on the details of the cosmological time parametrization.

\subsubsection{$dQ_{\TextH}/dt \neq 0$  ({\it i.e.}, $k_{\TB} T_{\BH} \geq m_{e} c^2$)}
\label{subsubsec:dQneq0} 
As the subcase of $T_{\BH} > T_{\env}$, we consider the Hawking radiation which the BH is emitting charged particles ({\it i.e.,} $d Q_{H}/dt \neq 0$). This happens when $M_{\BH} \leq 10^{-17} M_{\odot}$. In this case, the entropy change of a BH given in Eq.~\eqref{dSBHdt} becomes
\begin{align}
\frac{d S_{\BH}}{dt} = \frac{2 \pi k_{\TB 0} \tG_0 \tM_0^2}{\thbar_0 \tc_0} \frac{1 + \sqrt{1 - B(a)}}{\sqrt{1 - B(a)}} \left[ \left( 1 + \sqrt{1 - B(a)} \right) \frac{1}{M} \frac{dM_{\TextH}}{dt} - B(a) \frac{1}{Q} \frac{dQ_{\TextH}}{dt} + \frac{b}{8} H B(a) \right] \label{dSBHdt2} \,.
\end{align}

It is natural that a charged BH preferentially emits charged particles of the same sign as its own charge at a rate that depends on $Q$
\begin{align}
	\frac{d Q_{H}}{d t} = - \left( \frac{e |Q|}{Q} \right) \frac{dN_{H}}{dt} \label{dQdt} \,,
\end{align} 
where $e = -1.6 \times 10^{-19} C$ is the electron charge and $e dN_{H}/dt$ is the net emission rate of charge. Then the total emission rate of all particles from the BH is given by 
\begin{align}
\frac{d N_{\Tot}}{dt} = - \frac{\tc_0^2}{E_{\av 0}} \frac{d M_{H}}{dt} \geq \frac{dN_{H}}{dt} \label{dNTotdt} \,,
\end{align}
where $E_{\av 0} \approx 5 k_{\TB} T_{\BH 0} = 5 \thbar_0 \tc_0^3 /(2 \pi \tG_0 \tM_0)$ is the average energy of particles emitted by the BH \cite{Page:1976prd}. By combining Eqs.~\eqref{dQdt} and \eqref{dNTotdt}, one obtains 
\begin{align}
\left| \frac{B}{Q} \frac{d Q_{H}}{d t} \right| \leq \frac{\tc_0^2}{E_{\av}} \left|  \frac{e B}{Q} \frac{d M_{H}}{dt} \right| \approx  \frac{\left|  \te_0 \tQ_0 \right| }{10 \tepsilon_0 \thbar_0 \tc_0 \tM_0}  \frac{\left(1 + \sqrt{1-B} \right)^2}{\sqrt{1-B}} \frac{d M_{H}}{dt} \label{QdQHdt} \,. 
\end{align}
Thus, the second term of the entropy change given in Eq.~\eqref{dSBHdt2} lies in the range
\begin{align}
0 \leq \frac{d S_{\BH}}{dt} \Bigl|_{\mathrm{II}} \leq -\frac{2 \pi k_{\TB 0} \tG_0 \tM_0 \left|  \te_0 \tQ_0 \right| }{5 \tepsilon_0 \left( \thbar_0 \tc_0 \right)^2}  \frac{\left(1 + \sqrt{1-B} \right)^3}{1-B} \frac{d M_{H}}{dt} \equiv \frac{d S_{\BH}}{dt} \Bigl|_{\mathrm{II}}^{(\text{up})} \label{dSBHdtIIup} \,.
\end{align}
If we adopt the mass loss due to the Hawking radiation in Eq.~\eqref{dMdt} into the above equation, then we can estimate the upper limit of this entropy change as
\begin{align}
 \frac{d S_{\BH}}{dt} \Bigl|_{\mathrm{II}}^{(\text{up})} = \frac{2 \pi \tk_{\TB 0} \tc_0^2 |\te_0 \tQ_0|}{5 \tepsilon_0 \thbar_0 \tG_0 \tM_0} \frac{\left( 1-B \right) a^{-b/4}}{240 \pi \left( 1+ \sqrt{ 1 -B }  \right)^3}= 2.02 \left( \frac{M_{\odot}}{\tM_0} \right)  \left( \frac{\tQ_0}{C} \right) \frac{\left( 1-B \right) a^{-b/4}}{240 \pi \left( 1+ \sqrt{1 -B} \right)^3} \left[ \frac{J}{K \cdot s} \right] \,. \label{dSBHdtIIupnum}
\end{align}
Now we can compare the magnitudes of all three contributions of the BH entropy and these are shown in Table~\ref{tab:SBHcaseII}. The first contribution of the BH entropy change ($10^{17}$) is dominant for the MBH. The second contributions of the change of the BH entropy ($10^{-2}$) are dominant for both YBH and the specific BH. Thus, the sign of $b$ is irrelevant to satisfy the generalized second law of the BH thermodynamics. 

\begin{table}[htbp]
	\centering
\begin{tabular}{|c|c|c|c|}
	\hline 
	& MBH  & YBH & specific BH \\
	\hline
	M [$M_{\odot}$] & $10^{-38}$ & $10^{-19}$ & $10^{-17} $  \\
	\hline
	$Q_{\Teq}$ [$C$] & $10^{-36}$ & $10^{-17}$ & $10^{-15}$  \\
	\hline
	$dS_{\BH}/dt|_{\mathrm{I}} $ [$J/(K\cdot s)$] & $10^{17}$ & $10^{-2}$ & $10^{-4}$  \\
	\hline
	$dS_{\BH}/dt|_{\mathrm{II}}^{(\text{eq})} $ [$J/(K\cdot s)$] & $10^{-2}$ & $10^{-2}$ & $10^{-2}$  \\
	\hline
	$dS_{\BH}/dt|_{\mathrm{III}}^{(\text{eq})}$ [$J/(K\cdot s)$] & $10^{-72} (h\cdot b)$ & $10^{-48} (h\cdot b)$ & $10^{-40} (h\cdot b)$\\ 
	\hline
	$dS_{\text{tot}}^{(\text{eq})}/dt $ [$J/(K\cdot s)$] & $10^{17}$ & $10^{-2}$ & $10^{-2}$  \\
	\hline
\end{tabular}
\caption{The contributions of the change of BH entropy for the different BHs. The superscript (eq) denotes when one adopts the equal charge of BH in the calculation. }
\label{tab:SBHcaseII}
\end{table}

\subsection{$\tT_{\BH} \leq T_{\env}$}
\label{subsec:TBHleqTenv} 
If the temperature of BH is equal to or less than the temperature of its environment, then the BH accretes from its surroundings faster than it Hawking radiates. This accretion increases the mass of the BH $M$, further lowering its temperature, $T_{\BH}$. During accretion, the amount of the increase of the BH mass-energy ($M c^2$) should be equal to that of the decrease of the energy of the environment, $E_{\env}$.  The thermodynamical definitions of the temperature of an object is related to its entropy as 
\begin{align}
\beta_{i} \equiv \left( k_{\TB} T_{i} \right)^{-1} \equiv \frac{\partial S_{i}}{\partial E_i} = \frac{1}{c^2} \frac{\partial S_i}{\partial M_i} \Bigl|_{Q_i } \label{betai} \,.
\end{align}
Thus, if the temperature of the environment is greater than that of the BH, $\tT_{\BH} \leq T_{\env}$, then the increase in BH entropy due to accretion, $S_{\BH}$, must be greater than the decrease in the entropy of its environment, $S_{\env}$ as shown in Eq.~\eqref{betai}. We can also consider both the increase and the decrease of BH entropies due to accretion and Hawking radiation. For this consideration, we regard a cold large BH in a warm thermal bath which absorbs energy and emits radiations at a rate
\begin{align}
\frac{dE}{dt} \Bigl|_{\text{abs}} = \frac{\pi^2 \sigma_{\text{S}}}{60 \hbar^3 \tc^2} \left( \tk_{\TB} T_{\env} \right)^4 \quad , \quad
\frac{dE}{dt} \Bigl|_{\text{emt}} =  \frac{\pi^2 \sigma_{\text{S}}}{60 \hbar^3 \tc^2} \left( \tk_{\TB} T_{\BH} \right)^4  \,, \text{where} \quad \sigma_{\text{S}} = \frac{27 \pi \tG_0^2 \tM_0^2}{\tc_0^4} \label{dEdtemt} \,. 
\end{align} 
The increase in $S_{\BH}$ due to accretion ($dE/dt |_{\text{abs}}$) is greater than the decrease in $S_{\BH}$ due to Hawking radiation ($dE/dt |_{\text{emt}}$). Thus, the generalized entropy of the system is increased $\Delta S \geq 0$ in this case irrelevant to the sign of $b$.

\end{document}